\documentclass[copyright,creativecommons]{eptcs}
\providecommand{\event}{ICLP 2021} 
\usepackage{underscore}            

\usepackage{algorithm}
\usepackage{algorithmic}
\usepackage{amsmath}
\usepackage{booktabs}
\usepackage[T1]{fontenc}
\usepackage{graphicx}
\usepackage[utf8]{inputenc}
\usepackage{caption}
\usepackage{soul}
\usepackage{url}
\usepackage{xcolor}
\usepackage{multirow}

\def\naf{not \:\: }

\title{DMAPF: A Decentralized and Distributed Solver for Multi-Agent Path Finding Problem with Obstacles}
\author{Poom Pianpak \qquad\qquad Tran Cao Son
\institute{New Mexico State University\\
Las Cruces, New Mexico, USA}
\email{\{ppianpak,tson\}@cs.nmsu.edu}}

\def\titlerunning{DMAPF: A Decentralized and Distributed Solver for MAPF Problem with Obstacles}
\def\authorrunning{Poom Pianpak and Tran Cao Son}
\begin{document}
\maketitle

\begin{abstract}
\emph{Multi-Agent Path Finding} (MAPF) is a problem of finding a sequence of movements for agents to reach their assigned location without collision. 
Centralized algorithms usually give optimal solutions, but have difficulties to scale without employing various techniques -- usually with a sacrifice of optimality; but solving MAPF problems with the number of agents greater than a thousand remains a challenge nevertheless. 
To tackle the scalability issue, we present \emph{DMAPF} -- a decentralized and distributed MAPF solver, which is a continuation of our recently published work, \emph{ros-dmapf}~\cite{pianpak2019distributed}. 
We address the issues of \emph{ros-dmapf} where it (\textit{i})~only works in maps without obstacles; and (\textit{ii})~has a low success rate with dense maps. 
Given a MAPF problem, both \emph{ros-dmapf} and \emph{DMAPF} divide the map spatially into subproblems, but the latter further divides each subproblem into disconnected regions called areas. 
Each subproblem is assigned to a distributed solver, which then individually creates an abstract plan -- a sequence of areas that an agent needs to visit -- for each agent in it, and interleaves agent migration with movement planning.   
\emph{Answer Set Programming}, which is known for its performance in small but complex problems, is used in many parts including problem division, abstract planning, border assignment for the migration, and movement planning.  
\emph{Robot Operating System} is used to facilitate communication between the solvers and to enable the opportunity to integrate with robotic systems. 
\emph{DMAPF} introduces a new interaction protocol between the solvers, and mechanisms that together result in a higher success rate and better solution quality without sacrificing much of the performance. 
We implement and experimentally validate \emph{DMAPF} by comparing it with other state-of-the-art MAPF solvers and the results show that our system achieves better scalability. 
\end{abstract}

\section{Introduction}

Robots have been making their way into human life. 
From household robots to self-driving cars and industrial robots, it is expected that the number of robots will keep increasing in the future. 
To cope with such growth, robot controlling systems need to be designed with scalability in mind. 
We took an inspiration from autonomous warehouse systems\footnote{\url{https://amazonrobotics.com}}\textsuperscript{,}\footnote{\url{https://locusrobotics.com}} where the retrieval and storage tasks are done autonomously by mobile robots. 
When the system receives an order, it assigns a set of robots to retrieve shelves containing the products to a human operator 
for fulfillment of the order, then store the shelves back in their appropriate place. 
%


In this paper, we only consider the problem of finding a plan for agents to reach their assigned location (i.e., goal) without collision. 
The problem is called \emph{Multi-Agent Path Finding} (MAPF) and will be formally introduced in Section~\ref{sec:mapf}. 
Most existing MAPF algorithms are centralized (i.e., having a central unit overseeing the entire solving process) such as \emph{WHCA*}~\cite{silver2005cooperative}, \emph{asprilo}~\cite{gebser2018experimenting}, and \emph{CBS}~\cite{sharon2015conflict}. 
While there are attempts to improve the scalability of centralized algorithms using various techniques, such as abstraction (e.g., planning for each agent independently then combine the partial plans to obtain the solution -- while resolving any conflict), dealing with MAPF problems with a large number of agents (greater than a thousand) still remains a challenge. 
We have designed a decentralized and distributed MAPF algorithm, named \emph{DMAPF} -- \textbf{D}istributed \textbf{M}ulti-\textbf{A}gent \textbf{P}ath\textbf{f}inder -- with scalability in mind. 
\emph{DMAPF} is able to take advantage of distributed computing to cope with the possibility of having an ever-increasing problem size. 
An input problem to \emph{DMAPF} is encoded in \emph{answer set program} as generated by the \emph{ASPRILO} project~\cite{gebser2018experimenting} and solved using mainly the \emph{answer set programming}, which will be introduced in Section~\ref{sec:asp}. 
The communication between distributed components in \emph{DMAPF} is facilitated by the \emph{Robot Operating System} (ROS), which will be introduced in Section~\ref{sec:ros}. 

\emph{DMAPF} is an improvement over our original system, \emph{ros-dmapf}~\cite{pianpak2019distributed}. 
We address the issues in \emph{ros-dmapf} where it (\textit{i})~only works in maps without obstacles; and (\textit{ii})~has a low success rate with dense maps. 
\emph{DMAPF} shares the same idea with \emph{ros-dmapf} in that, given a MAPF problem, it divides the problem spatially and assigns each divided subproblem to a distributed solver, which is a ROS node. 
The difference is that \emph{DMAPF} also further divides each subproblem into disconnected regions called areas. 
This, together with a few other changes, allow \emph{DMAPF} to deal with having obstacles. 
The details of problem division will be explained in Secion~\ref{sec:problem-division}. 
After the subproblems have been distributed, each solver individually creates an abstract plan -- a sequence of areas that an agent needs to visit to reach the area that contains its goal -- for each agent in the given subproblem. 
The details of abstract plan creation will be explained in Section~\ref{sec:problem-solving-abs}. 
After the plans have been made for every agent, the solvers interleave communicating with neighboring solvers to send/receive migrating agents (details in Sections~\ref{sec:problem-solving-comm-1},~\ref{sec:problem-solving-comm-2}, and~\ref{sec:problem-solving-comm-3}) with movement planning (details in Section~\ref{sec:problem-solving-move}) along each round. 
This differs from \emph{ros-dmapf} where it does the communication from beginning to end first, then finds a movement plan for each round in a single attempt. 
If \emph{ros-dmapf} is unable to make a plan during the movement planning phase, it would be difficult to adjust the border assignment since assignments for the next rounds have already been agreed upon, whereas \emph{DMAPF} can easily adjust the assignment and retry. 
This, together with changes in migration protocol, allow \emph{DMAPF} to achieve a higher success rate and better solution quality than \emph{ros-dmapf}.  
\emph{DMAPF} has been implemented and compared with other state-of-the-art solvers.
Section~\ref{sec:experiments} shows results of the comparison.  Section~\ref{sec:conclusion} concludes with discussion on the results, advantages, limitations, and future work. 

\section{Background}
\subsection{Multi-Agent Path Finding Problem}
\label{sec:mapf}
A \emph{Multi-Agent Path Finding} (MAPF) problem can be defined as a quadruple $P = (G, A, S, T)$, where $G = (V,E)$ is a graph such that $V$ is a set of vertices corresponding to locations in the graph, and $E \subseteq V \times V$ denotes edges between two locations; $A$ is a set of agents; and $S \subseteq A \times V$ and $T \subseteq A \times V$ denote start and goal locations of the agents, respectively. 

Agents can move from $v_1$ to $v_2$ where $v_1, v_2 \in V$ if $(v_1, v_2) \in E$, under the restrictions: (\textit{a})~two agents cannot swap locations in a single time step; and (\textit{b})~each location can be occupied by at most one agent at a time. 
A path for an agent $a$ is a sequence of vertices $\alpha_a = \langle v_1,\ldots,v_n \rangle$ if (\textit{i})~agent $a$ starts at $v_1$ (i.e.,~$(a,v_1) \in S$); and (\textit{ii})~there is an edge between any two subsequent vertices $v_i$ and $v_{i+1}$ (i.e.,~$(v_i,v_{i+1}) \in E$), or they are the same vertex (i.e.,~$v_{i+1} = v_{i}$). 
An agent $a$ completes its order $T_a = \{v \mid (a,v) \in T\}$ via a path $\alpha_a = \langle v_1,\ldots,v_n \rangle$ if $T_a \subseteq \{ v_1,\ldots,v_n\}$. 
A \emph{solution} of a MAPF problem $P$ is a collection of paths $Sol = \{\alpha_a \mid a \in A\}$ such that all orders in $T$ are completed.


In our work, we assume that (\textit{i})~each agent has a different start location (i.e.,~$\forall_{a}\exists_{v}((a \in A \to (a,v) \in S) \land \neg \exists_{\hat{v} \neq v}((a,\hat{v}) \in S)) \land \neg \exists_{\hat{a} \neq a} ((\hat{a},v) \in S))$); (\textit{ii})~each agent either has no goal or has a distinct goal location (i.e.,~$\forall_{a_1,a_2,v}((a_1,v),(a_2,v) \in T \to a_1 = a_2)$); (\textit{iii})~each agent is at its goal at the last time step (i.e.,~$\forall_{a,v}(((a,v) \in T \land \alpha_a = \langle v_1,\ldots,v_n \rangle) \to v = v_n)$); and (\textit{iv})~the graph is grid-based. 
These assumptions are common among most multi-agent path finding solvers.

\subsection{Answer Set Programming}
\label{sec:asp}
Let us provide some general background on \emph{Answer Set Programming} (ASP). 
Consider a logic language $L=\langle C, P, V \rangle$, where $C,P,V$ are sets of constants, predicate symbols, and variables, respectively, and the notions of terms, atoms, and literals are traditional. 

An \emph{answer set program} $\Pi$ is a set of \emph{rules} of the form
\begin{equation} \label{lprule1}
c \leftarrow a_1,\ldots,a_m, \naf b_1,\ldots, \naf b_n 
\end{equation}
Each $a_i$ and $b_i$ is a literal from $L$, and each $\naf b_i$ is called a negation-as-failure literal (or naf-literal). 
$c$ can be a literal or omitted. 
A program is a {\em positive program} if it does not contain naf-literals. 
A \emph{non-ground rule} is a rule that contains variables; otherwise, it is called a \emph{ground rule}. 
A rule with variables is simply used as a shorthand for the set of its ground instances from the language $L$. 
If $n = m = 0$, then the rule is called a \emph{fact}. If $c$ is omitted, then the rule is called an \emph{ASP constraint}.  
 
A set of ground literals $X$ is \emph{consistent} if there is no atom $a$ such that $\{a, \neg a\} \subseteq X$. 
A literal $l$ is true (resp. false) in a set of literals $X$ if $l \in X$ (resp. $l \not\in X$). 
A set of ground literals $X$ satisfies a ground rule of the form~\eqref{lprule1} if either of the following is true: (\textit{i})~$c \in X$; (\textit{ii})~some  $a_i$ is false in $X$; or (\textit{iii})~some  $b_i$ is true in $X$. 
A solution of a program, called an \emph{answer set}~\cite{GelfondL90}, is a consistent set of ground literals satisfying the following conditions: 

\begin{itemize}
\item If $\Pi$ is a \emph{ground program} (i.e., a program whose rules are all ground), then its answer set $S$ is defined by the following:

\begin{list}{$\circ$}{\topsep=1pt \parsep=0pt \itemsep=1pt}
\item If $\Pi$ does not contain any naf-literals, then $S$ is an answer set of $\Pi$ if it is a consistent and 
subset-minimal set of ground literals satisfying all rules in $\Pi$. 

\item If $\Pi$ contains some naf-literals, then $S$ is an answer set of $\Pi$ if it is an answer set of the \emph{program reduct} $\Pi^S$. $\Pi^S$ is obtained from  $\Pi$ by deleting (\textit{i})~each rule that has a naf-literal \emph{not} $b$ in its body with $b \in S$; and (\textit{ii})~all naf-literals in the bodies of the remaining rules. 
\end{list}

\item If $\Pi$ is a \emph{non-ground program} (i.e., a program whose rules include non-ground rules), then $S$ is an answer set of $\Pi$ if it is an answer set of the program consisting of all ground instantiations of the rules in $\Pi$.
\end{itemize}

The ASP language includes also language-level extensions to facilitate the encoding of aggregates ($min$, $max$, $sum$, etc.), range specification of variables, and allowing choice of literals. 
In ASP, one solves a problem by encoding it as an ASP program whose answer sets correspond one-to-one to the  solutions of the problem~\cite{MarekT99,Niemela99}. Answer sets of ASP programs can be computed using ASP solvers like \textsc{clasp}~\cite{GebserKNS07} and \textsc{dlv}~\cite{citrigno1997dlv}. 

Early ASP rests upon a single-shot approach to problem solving -- an ASP solver takes a logic program, computes its answer sets, and exits. 
Unlike this, recently developed multi-shot ASP solvers provide operative solving processes for dealing with continuously changing logic programs. 
For controlling such solving processes, the declarative approach of ASP is combined with imperative means. 
In \textit{clingo} \cite{gekakasc14b}, this is done by augmenting an ASP encoding with C or Python procedures. 
The instrumentation includes methods for adding/grounding rules, setting truth values of (external) atoms, computing the answer sets of current program, etc.


\subsection{Robot Operating System}
\label{sec:ros}
The \emph{Robot Operating System} (ROS) is an open-source framework designed for building robotics systems which is distributed in nature~\cite{quigley2009ros}. 
We adopt ROS because of its scalability and support for the development of heterogeneous clusters of software.
ROS provides client libraries\footnote{Visit~\url{http://wiki.ros.org/Client\%20Libraries}~for a full list of ROS client libraries} for software written in different languages (e.g., C++, Python, Lisp) to communicate. 



A ROS system must consist of a \emph{roscore} at a bare minimum, and may consist of multiple \emph{ROS nodes}. 
A roscore is a set of prerequisites to run a ROS-based system, and it consists of a \emph{ROS master} among a few other things. 
A ROS node is an individual process that does some computation. 
ROS nodes working together for a particular task may be organized as a \emph{package}, and they can be on different networks. 
For a node $n_1$ to communicate with another node $n_2$, $n_1$ first needs to locate $n_2$ via the ROS master, then $n_1$ can communicate directly with $n_2$ as a peer-to-peer network. 
There are mainly two forms of communication between ROS nodes: 

\begin{enumerate}
\item \emph{Publish-Subscribe} -- nodes are connected via a \emph{topic}, which is a named bus. 
A node sending (resp. listening to) messages on a topic is called a \emph{publisher} (resp. \emph{subscriber}). 
One node can both be a publisher and a subscriber on the same or multiple topics.
A topic may have zero or more publishers and/or subscribers. 
\item \emph{Request-Response} -- two nodes follow an RPC interaction through a \emph{service}. 
A node that provides (resp. calls) a service is called a \emph{service server} (resp. \emph{service client}). 
There can only be one service server, but possibly multiple service clients for a single service. 
Calls from the service clients will be put into a queue and processed one-by-one. 
\end{enumerate}

%

\section{Methodology}
\label{sec:methodology}

Algorithm~\ref{alg:dmapf} shows an overview of \emph{DMAPF}. 
The algorithm takes a MAPF problem $P$ and the dimension $dx \times dy$ of desired subproblems as inputs, and produces the solution if it could find one.
Line~\ref{alg:overview-divide} divides the given problem $P$ into smaller subproblems: $P_1, \dots, P_n$, and provides the $Links$ information telling which pairs of \emph{areas} within the subproblems are \emph{connected} (i.e., there exists some border between them). 
This step will be explained in Section~\ref{sec:problem-division}. 
The problem can also be divided by hand as \emph{DMAPF} does not put restrictions on how the subproblems have to be divided, i.e., they do not have to be rectangles of dimension $dx \times dy$. 
Lines~\ref{alg:overview-solve-start}-\ref{alg:overview-solve-end} create solvers $s_1,\dots,s_n$ as \emph{ROS nodes} with the divided subproblem $P_i$ as their input. 
How the solvers work together, which is the core of the algorithm, will be explained in Section~\ref{sec:problem-solving}. 
Solver $s_1$ has an extra responsibility, besides solving its own subproblem, to aggregate partial plans from all the solvers and create the solution as an output. 
Therefore, line~\ref{alg:overview-wait} waits for $s_1$ to finish its own task of solving $P_1$, combine partial plans from all the solvers, and produce the solution at line~\ref{alg:overview-return}. 

\begin{algorithm}
\caption{\emph{DMAPF}}
\label{alg:dmapf}
\textbf{Input}: $P$, $dx$, $dy$\\
\textbf{Output}: A solution or none
\begin{algorithmic}[1] 
\STATE $(P = \{P_1, \dots, P_n\}, Links) \leftarrow Divide(P, dx, dy)$ \label{alg:overview-divide}
\FOR {\textbf{each} $P_i \in P $} \label{alg:overview-solve-start}
\STATE Create solver $s_i$ for solving $P_i$ \COMMENT{in parallel} \label{alg:overview-solve}
\ENDFOR \label{alg:overview-solve-end}
\STATE Wait for $s_1$ \label{alg:overview-wait}
\RETURN $s_1.solution$ \label{alg:overview-return}
\end{algorithmic}
\end{algorithm}

\begin{figure}
\centering
\includegraphics[width=.6\textwidth]{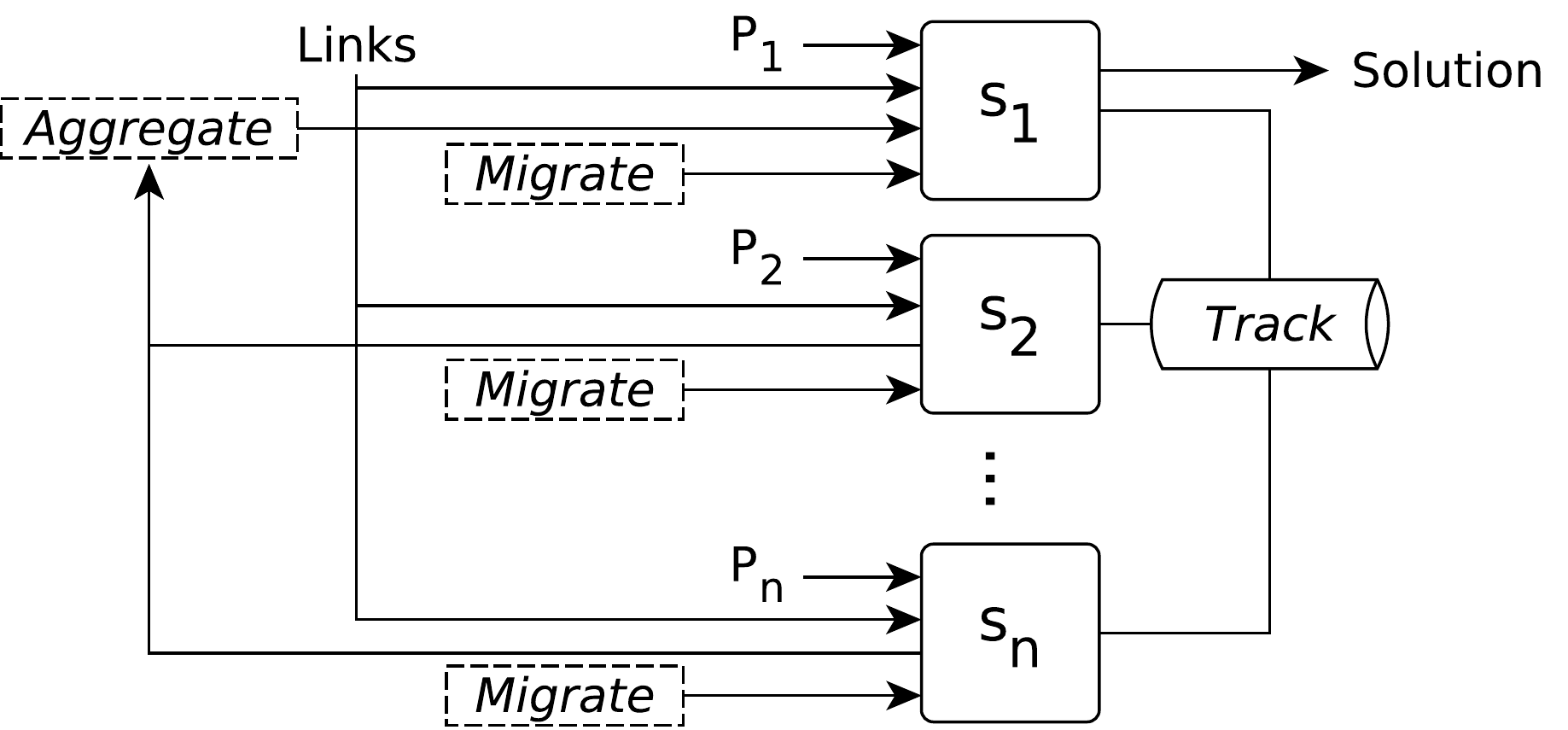}
\caption{
An overview of a \emph{DMAPF} system. 
}
\label{fig:solve-overview}  
\end{figure}

Figure~\ref{fig:solve-overview} depicts an overview of a \emph{DMAPF} system. 
In there, the problem has been divided into subproblems $P_1, P_2, \dots, P_n$ and each of them is an input to solvers $s_1, s_2, \dots, s_n$, respectively. 
All solvers have \emph{Links} information telling which pairs of them are \emph{connected}. 
They use this information as an abstract map to find abstract plans (details in Section~\ref{sec:problem-solving-abs}). 
All solvers connect to \emph{Track} topic to synchronize and exchange information to determine whether they have finished their work, or if not, which of them still have work to do. 
Each of them also provides a \emph{Migrate} service for connected solvers to call if there is any robot that needs to move between them. 
Solver $s_1$ provides \emph{Aggregate} service for the other solvers to submit their plan, which constitutes a part of the solution, then it will combine them to produce the final solution.

\subsection{Problem Division}
\label{sec:problem-division}

Figure~\ref{fig:prog-divide} shows a set of programs used in problem division. 
The uses of these programs are controlled by a C++ implementation. 
First, given a MAPF problem $P$, as generated by the \emph{ASPRILO} project~\cite{gebser2018experimenting}, and the dimension $dx \times dy$ of desired subproblems, program \verb|base| is called to determine the following atoms: 

\begin{itemize}
\item $d(D,(X,Y))$ -- direction $(X,Y)$ that a robot can move in each time step. 
$D$ is a number representing the direction, which will be later used in movement planning (Section~\ref{sec:problem-solving-move}). 
In our setting, the robots can move one node at a time to either left, right, up, or down, which correspond to $(-1,0)$, $(1,0)$, $(0,-1)$, or $(0,1)$, respectively. 
\item $goal(R,N)$ -- robot $R$ has its goal at node $N$. 
To restrict each robot to have at most one distinct goal, we assume that there is \textbf{at most} one order per robot, where the order is to pick a distinct product $P$ on a distinct shelf $S$, which is at node $N$ at coordinate $C$. 
\end{itemize}

\begin{figure}
\begin{verbatim}
#program base.
d(0,(-1,0)). d(1,(1,0)). d(2,(0,-1)). d(3,(0,1)).
goal(R,N) :- init(object(order,R),value(line,(P,_))),
             init(object(product,P),value(on,(S,_))),
             init(object(shelf,S),value(at,C)), 
             init(object(node,N),value(at,C)).

#program divide(s, x_min, x_max, y_min, y_max).
node(s,N,(X,Y)) :- init(object(node,N),value(at,(X,Y))),
                   X >= x_min, X < x_max, Y >= y_min, Y < y_max.
robot(s,R,N) :- init(object(robot,R),value(at,C)), node(s,N,C).
edge(s,N1,N2) :- node(s,N1,(X,Y)), node(s,N2,(X+DX,Y+DY)), d(_,(DX,DY)).
edge(s,N,N) :- node(s,N,_).
border(s,N) :- node(s,N,(x_min,Y)). border(s,N) :- node(s,N,(x_max-1,Y)).
border(s,N) :- node(s,N,(X,y_min)). border(s,N) :- node(s,N,(X,y_max-1)).

#program link.
link(S1,S2,N1,N2) :- border(S1,N1), border(S2,N2), node(S1,N1,(X,Y)),
                     node(S2,N2,(X+DX,Y+DY)), S1 != S2, d(_,(DX,DY)).

#program result.
i(N,C) :- node(S,N,C), area(S,A,N).
o(N2,C) :- link(S1,S2,N1,N2), area(S1,A,N1), node(S2,N2,C).
x(N1,N2,D) :- area(S,A,N1), area(S,A,N2), node(S,N1,(X1,Y1)),
              node(S,N2,(X2,Y2)), d(D,(X2-X1,Y2-Y1)).
x(N2,N1,D) :- link(S1,S2,N1,N2), area(S1,A,N1), node(S1,N1,(X1,Y1)),
              node(S2,N2,(X2,Y2)), d(D,(X1-X2,Y1-Y2)).
l(N1,C1,N2,C2) :- link(S1,S2,N1,N2), area(S1,A1,N1), node(S1,N1,C1),
                  area(S2,A2,N2), node(S2,N2,C2).
l(A1,A2) :- link(S1,S2,N1,N2), area(S1,A1,N1), area(S2,A2,N2).
\end{verbatim}
\caption{The collection of programs used for problem division.}
\label{fig:prog-divide}
\end{figure}

Subproblems are determined in an iterative manner. 
Suppose $x_{min}$, $x_{max}$, $y_{min}$, and $y_{max}$ are the left, right, top, and bottom-most points of a given problem $P$, there would be a total of $\lceil \frac{x_{max} - x_{min} + 1}{dx} \rceil \cdot \lceil \frac{y_{max} - y_{min} + 1}{dy} \rceil$ subproblems (if the map is rectangular). 
The map of the top-left subproblem would be bounded within the range of $x = [x_{min},~x_{min} + dx)$ and $y = [y_{min},~y_{min} + dy)$. 
The boundaries of the other subproblems can be determined by adjusting $x_{min}$, $x_{max}$ (resp. $y_{min}$, $y_{max}$) by $dx$ (resp. $dy$). 
Program \verb|divide| is repeatedly called (with external atoms which we omit) to determine the information pertaining to each subproblem $S$ as follows: 

\begin{itemize}
\item $node(S,N,C)$ -- node $N$ is in subproblem $S$ at coordinate $C$. 
\item $robot(S,R,N)$ -- robot $R$ starts at node $N$ which is in subproblem $S$. 
\item $edge(S,N_1,N_2)$ -- nodes $N_1$ and $N_2$ are next to each other in subproblem $S$. 
\item $border(S,N)$ -- node $N$ is a border in subproblem $S$. 
\end{itemize}

Program \verb|link| is then called after all the subproblems have been created from the calls to program \verb|divide|. 
It determines the following atom:

\begin{itemize}
\item $link(S_1,S_2,N_1,N_2)$ -- node $N_1$ in subproblem $S_1$ is next to node $N_2$ in subproblem $S_2$. 
Solvers of $S_1$ and $S_2$ are considered to be \emph{neighbors} if there is a link between them. 
\end{itemize}

Atoms $edge/3$ acquired from the calls to program \verb|divide| are used to determined areas within each subproblem.
An \emph{area} is a region where the nodes inside are all connected. 
Atoms $area(S,A,N)$ are added to denote that node $N$ is a part of area $A$ which is in subproblem $S$. 
Then, program \verb|result| is repeatedly called (with external atoms which we omit) to determine a set of information pertaining to each area $A$ as follows: 

\begin{itemize}
\item $i(N,C)$ -- node $N$ at coordinate $C$ is \textbf{i}n area $A$. 
\item $o(N,C)$ -- node $N$ at coordinate $C$ is right \textbf{o}utside of area $A$ (only one \emph{direction} away). 
\item $x(N_1,N_2,D)$ -- node $N_1$ is ne\textbf{x}t to node $N_2$, and they are in area $A$. 
$N_1$ can be a node in either $i/2$ or $o/2$, but $N_2$ can only be a node in $i/2$. 
\item $l(N_1,C_1,N_2,C_2)$ -- node $N_1$ at coordinate $C_1$ in area $A$ is next to node $N_2$ at coordinate $C_2$ in another area. 
\end{itemize}

Each subproblem $P_i$ consists of information about its areas, which includes the atoms acquired from the call to program \verb|result|. 
Besides the atoms, a subproblem also contains auxiliary information such as the number of nodes in different areas (within the subproblem), a list of its neighboring areas (and their solvers), and a list of nodes that are \emph{corners} (i.e., a node that connects to multiple areas) and the areas which they connect to. 
A set of atoms $l(A_1,A_2)$, which denotes that areas $A_1$ and $A_2$ are connected, is the content of $Links$ (in figure~\ref{fig:solve-overview}) which is an input to every solver.

\subsection{Problem Solving}
\label{sec:problem-solving}

After a MAPF problem $P$ has been divided into subproblems $P_1, P_2, \dots, P_n$ as explained in Section~\ref{sec:problem-division}, a solver $s_i$ is created for each subproblem $P_i$. 
The solvers $s_1, s_2, \dots, s_n$ execute algorithm~\ref{alg:solve} in \underline{parallel} to solve their own subproblem $P_i$, and at the end $s_1$ will combine their partial plans to produce the solution of the original problem $P$. 
The algorithm requires a few explanations as follows:

\begin{enumerate}
\item The lines where the first word is underlined denote that they involve communication with other solvers. 
Lines~\ref{alg:solve-determine-task-1}~and~\ref{alg:solve-determine-task-2} publish a message to the \emph{Track} topic. 
Lines~\ref{alg:solve-neg}, \ref{alg:solve-rej}, and \ref{alg:solve-cfm} send a migration request to the \emph{Migrate} service of the solver that contains area $a$, for each $a \in Send$. 
Lines~\ref{alg:solve-neg-end}, \ref{alg:solve-rej-end}, and \ref{alg:solve-cfm-end} wait to receive a number of $Migrate$ service calls according to the number of areas in $Recv$. 

\item The global variables: $Active$, $Send$, and $Recv$ (defined at line~\ref{alg:solve-global-vars}), are determined from the communication between solvers. 
$Active$ keeps track of solvers that still have work to do, i.e., there is a movement plan to be made or there exists a robot that needs to migrate; and has the same content across all the solvers. 
$Send$ (resp. $Recv$) stores areas that $s_i$ needs to send (resp. receive) migration requests to (resp. from). 
The contents of the variables are determined at lines~\ref{alg:solve-determine-task-1}~and~\ref{alg:solve-determine-task-2} by having each solver publishes whether it still has any robot that has not reached its goal (determines $Active$), and if so, which area they would need to leave and enter if the current area does not have their goal (determines $Send$ and $Recv$). 
Every solver (and area) can be uniquely identified by a number (i.e., an ID). 
A solver (resp. area) with an ID $i$ is denoted as $s_i$ (resp. $a_i$). 
For every pair of solvers $s_i$ and $s_j$, if $i < j$, then all areas in $s_i$ also have smaller IDs than those in $s_j$. 
When a solver $s_i$ wants to send a service request to another solver $s_j$, it can only do so if $i < j$ -- this does not mean that $s_j$ cannot send information to $s_i$, as it can do so in the response message -- RPC style. 
This design ensures that there is only one direction of service calls between each pair of neighboring solvers, thus reducing the overhead in communication by half. 

\item Function $A$ (defined at line~\ref{alg:solve-areas}) maps a round (denoted as a number starting from $0$) to a set of area instances. 
An \emph{area instance} (or \emph{instance}, for short) consists of necessary information about everything in the area such as nodes, corners, border assignment constraints, current robots ($R$), outgoing robots ($R_i$), incoming robots ($R_o$), and a movement plan. 
An instance can be constructed either from either (\textit{i})~an \emph{area}, as obtained after the problem division (line~\ref{alg:solve-construct-area-1} and~\ref{alg:solve-construct-missing-area-1}); or (\textit{ii})~another instance (lines~\ref{alg:solve-construct-area-2} and~\ref{alg:solve-construct-missing-area-2}). 
When an instance $\alpha_2$ is constructed from another instance $\alpha_1$, the information about the robots in $\alpha_2$ is determined from $\alpha_1$ as $\alpha_2.R = (\alpha_1.R \setminus \alpha_1.R_o) \cup \alpha_1.R_i$.
The other information is straightforward to determine, so we omit the details. 
Lines~\ref{alg:solve-construct-missing-area-1} deals with the case where an instance $\alpha$ for area $a$ has not been constructed (because area $a$ contains no robot), but there are some robot in another area that wants to migrate to $a$ -- so $\alpha$ needs to be constructed. 
Line~\ref{alg:solve-construct-missing-area-2} deals with the case where an instance $\alpha_2$ needs to be constructed based on another instance $\alpha_1$ of the same area $a$ in the most recent previous round; the problem is, $\alpha_1$ may not have been constructed at all. 
If such case happens, $\alpha_2$ would be constructed from area $a$. 
\end{enumerate}

The algorithm revolves around the notions of (\textit{i})~abstract planning (line~\ref{alg:solve-abs}); (\textit{ii})~migration; and (\textit{iii})~movement planning (line~\ref{alg:solve-movement}). 
The migration process is divided into three steps: (\textit{i})~negotiation (lines~\ref{alg:solve-neg}-\ref{alg:solve-neg-end}); (\textit{ii})~rejection (lines~\ref{alg:solve-rej}-\ref{alg:solve-rej-end}); and (\textit{iii})~confirmation (lines~\ref{alg:solve-cfm}-\ref{alg:solve-cfm-end}). 
The details of each notion will be explained sequentially -- as how the algorithm works -- in the following subsections. 

\begin{algorithm}
\caption{$Solve$}
\label{alg:solve}
\textbf{Input}: $P_i$, $Links$\\
\textbf{Output}: A solution or none\\
\textbf{Parameters}: $s_i$ -- a number denoting this solver
\begin{algorithmic}[1] 
\STATE \textbf{global} $Active = Send = Recv = \{\}$ \label{alg:solve-global-vars}
\STATE $A = \{\}$ \label{alg:solve-areas}
\FOR {\textbf{each} area $a$ in $P_i$ that exists a robot}
  \STATE $A = A \cup \{(0, \alpha)\}$ where an instance $\alpha$ is constructed from $a$ \label{alg:solve-construct-area-1}
  \FOR {\textbf{each} robot $r$ in $\alpha$}
    \STATE Create an abstract plan for $r$ using $Links$ as an abstract map \label{alg:solve-abs}
  \ENDFOR
\ENDFOR
\STATE \underline{Determine} $Active$, $Send$, and $Recv$ from $A(0)$ of all the solvers \label{alg:solve-determine-task-1}
\STATE $A = A \cup \{(0, \alpha)\}$ where an instance $\alpha$ is constructed from an area $a$ (resp. $b$) that will be sending (resp. receiving) a request to (resp. from) the solver of area $c \in Send$ (resp. $d \in Recv$) -- where such an instance has not been created \label{alg:solve-construct-missing-area-1}

\STATE
\STATE $i = 0$ \label{alg:solve-round}
\WHILE {\TRUE}
\IF {$s_i \in Active$}
\STATE \underline{Send} a negotiation request to the solver of area $a$ \textbf{for each} $a \in Send$ \label{alg:solve-neg}
\STATE \underline{Wait} until $|Recv|$ negotiation requests have been received \label{alg:solve-neg-end}
\STATE \underline{Send} a rejection request to the solver of area $a$ \textbf{for each} $a \in Send$ \label{alg:solve-rej}
\STATE \underline{Wait} until $|Recv|$ rejection requests have been received \label{alg:solve-rej-end}
\STATE Make a movement plan \textbf{for each} instance in $A(i)$ \label{alg:solve-movement}
\STATE \underline{Send} a confirmation request to the solver of area $a$ \textbf{for each} $a \in Send$ \label{alg:solve-cfm}
\STATE \underline{Wait} until $|Recv|$ confirmation requests have been received \label{alg:solve-cfm-end}

\STATE $A = A \cup \{(i + 1, \alpha_{i+1})\}$ where an instance $\alpha_{i+1}$ is constructed from $\alpha_i \in A(i)$ \textbf{if} there exists a robot in $\alpha_i$ that still needs to move \label{alg:solve-construct-area-2}
\ELSE
\STATE \underline{Wait} for active solvers to finish their work
\ENDIF 

\STATE
\IF {$Active = \emptyset$}
\STATE \textbf{return} $solution \leftarrow$ aggregated partial plans from all the solvers \label{alg:solve-return}
\ELSE
\STATE \underline{Determine} $Active$, $Send$, and $Recv$ from $A(i + 1)$ of all the solvers \label{alg:solve-determine-task-2}

\STATE $A = A \cup \{(i + 1, \alpha_{i+1})\}$ where an instance $\alpha_{i+1}$ is constructed from an instance $\alpha_j$ of area $a$ (resp. $\beta_j$ of area $b$) for the most recent round $j$ where $\alpha_j$ (resp. $\beta_j$) $\in A(j)$, and $a$ (resp. $b$) will be sending (resp. receiving) a request to (resp. from) the solver of area $c \in Send$ (resp. $d \in Recv$) -- where such an instance (i.e., $\alpha_{i+1}$) has not been created; if such $j$ does not exist, then $\alpha_{i+1}$ is constructed from $a$ (resp. $b$) \label{alg:solve-construct-missing-area-2}

\STATE $i = i + 1$
\ENDIF
\ENDWHILE
\end{algorithmic}
\end{algorithm}

\subsubsection{Abstract Planning}
\label{sec:problem-solving-abs}
An \emph{abstract plan} is a sequence of connected areas that takes a robot from its initial area to reach the area that contains its goal. 
The program shown in figure~\ref{fig:prog-abstract} is used to make an abstract plan for a robot $r$.
$r(A,I)$ denotes that $r$ is in area $A$ at round $I$; and $g(A)$ denotes that the goal of $r$ is in area $A$. 
Program \verb|abstract(i)| is solved sequentially from $i = 0, \dots, h_a$ or until an answer set if found. 
If no answer set is found after $i = h_a$, then we consider there is no way for $r$ to reach its goal, and the algorithm terminates. 
$h_a$ is set to be the total number of areas across all the subproblems, so an abstract plan has to exist if it is possible for a robot to reach its goal (without considering the other robots). 

\begin{figure}[h]
\begin{verbatim}
#program abstract(i).
#external q(i).
1 { r(A2,i) : l(A1,A2) } 1 :- r(A1,i-1).
:- q(i), not r(A,i), g(A).
#show r/2.
\end{verbatim}
\caption{The program to make an abstract plan for a single robot.}
\label{fig:prog-abstract}
\end{figure}

\subsubsection{Negotiation}
\label{sec:problem-solving-comm-1}
\emph{Negotiation} is the first phase of the migration process. 
Its purpose is to find a set of distinct border assignments for migrating robots. 
The program shown in figure~\ref{fig:prog-migrate} is used to make such assignments.
It is designed to minimize \emph{makespan} -- the time step that all robots reach their goal. 
When a solver $s_1$ of an area $a_1$ sends a negotiation request to a neighboring solver $s_2$ of an area $a_2$, the message contains a set of robots that want to migrate, called $a_1.R_m$, with their current location, grouped and ordered by the number of steps left in their abstract plan to reach the last area (longest first); the result of the grouping will be referred to as \emph{tier}
After $s_2$ receives the request from $s_1$, it merges $a_1.R_m$ with $a_2.R_m$ and orders them by tier from high to low; the result of the merging will be referred to as $R_m$. 
It then computes two variables: the number of incoming (resp. outgoing) borders available, $n_{ai}$ (resp. $n_{ao}$). 
Let $n_l$, $n_{bi}$, and $n_{bo}$ denote the number of pairs of connected borders between $a_1$ and $a_2$, the number of incoming borders that have been blocked, and the number of outgoing borders that have been blocked, respectively, then $n_{ai} = n_l - n_{bi}$, and $n_{ao} = n_l - n_{bo}$. 
Then, $s_2$ goes through $R_m$ tier-by-tier. 
In each tier, if the robots are in $a_2$, then they are used to create either atoms $a/2$ or $b/2$; otherwise, they are used to create either atoms $c/2$ or $d/2$. 
Atoms $a/2$, $b/2$, $c/2$, and $d/2$ in program \verb|migrate| denote the current location of a robot. 
Whether to encode the robots in $a/2$ (resp. $c/2$) or $b/2$ (resp. $d/2)$ depends on the current number of robots that have been assigned out (resp. in), which we will refer to as $n_o$ (resp. $n_i$).
If the number of robots in the tier is less than or equal to $n_o$ (resp. $n_i$), then they are used to create $a/2$ (resp. $c/2$); otherwise, they are used to create $b/2$ (resp. $d/2)$. 
After the robots in the considering tier have been used to create the atoms, then the value of either $n_o$ or $n_i$ is increased by the number of robots added accordingly, and the tier is removed from $R_m$. 
$s_2$ will stop going though $R_m$ when either (\textit{i})~$R_m = \emptyset$; or (\textit{ii})~$n_o \geq n_{ao}$ and $n_i \geq n_{ai}$. 
The atom $l(L)$ in program \verb|migrate| denotes the \textbf{l}imit -- the number of robots expected to get the border assigned. 
$L$ could be calculated as $\min(\min(n_i, n_{ai}) + \min(n_o, n_{ao}), \max(n_{ai}, n_{ao}))$. 
A collection of atoms $m(R,N_1,N_2,D)$, where it denotes that robot $R$ can migrate from border $N_1$ in its current area to border $N_2$ in another area and the distance from its current node to $N_1$ is $D$, is obtained from program \verb|migrate|.
Each atom $m/4$ is checked whether $R$ is in area $a_1$ or $a_2$. 
$N_1$ if $R \in a_2$ (resp. $N_2$ if $R \in a_1$) is checked whether it is a \emph{corner} node (i.e., a node that connects to more than two areas); if so, a constraint \verb|:- m(_,N1,_,_).| (resp. \verb|:- m(_,_,N2,_).|) will be added, and the value of $n_{bo}$ (resp. $n_{bi}$) is increased by one. 
These constraints prevent $s_2$ from making border assignments for the next negotiation requests (within the same round) that would result in conflict at the corners. 
Finally, $s_2$ remembers the resulting border assignments and also send them to the caller, $s_1$. 


\begin{figure}[h]
\begin{verbatim}
#program migrate.
1 { m(R,N2,N3,|X1-X2|+|Y1-Y2|) : l(N2,(X2,Y2),N3,C3) } 1 :- a(R,(X1,Y1)).
0 { m(R,N2,N3,|X1-X2|+|Y1-Y2|) : l(N2,(X2,Y2),N3,C3) } 1 :- b(R,(X1,Y1)).
1 { m(R,N2,N3,|X1-X2|+|Y1-Y2|) : l(N3,C3,N2,(X2,Y2)) } 1 :- c(R,(X1,Y1)).
0 { m(R,N2,N3,|X1-X2|+|Y1-Y2|) : l(N3,C3,N2,(X2,Y2)) } 1 :- d(R,(X1,Y1)).
:- m(R1,N,_,_), m(R2,N,_,_), R1!=R2.
:- m(R1,_,N,_), m(R2,_,N,_), R1!=R2.
:- m(_,N1,N2,_), m(_,N2,N1,_).
:- C = #count { R : m(R,_,_,_) }, C!=L, l(L).
d(S) :- S = #sum { D : m(_,_,_,D) }.
#minimize { S : d(S) }.
#show m/4.
\end{verbatim}
\caption{The program to make border assignments for multiple robots.}
\label{fig:prog-migrate}
\end{figure}

\subsubsection{Rejection}
\label{sec:problem-solving-comm-2}
\emph{Rejection} is the second phase of the migration process. 
Its purpose is to prevent collision at corners that reside between areas with lower and higher IDs. 
Suppose there are three solvers: $s_1$, $s_2$, and $s_3$, which contain areas $a_1$, $a_2$, and $a_3$, respectively; two connected pairs of the areas: (\textit{1}) $a_1$ and $a_2$, and (\textit{2}) $a_2$ and $a_3$; and a corner $c$ in $a_2$ that connects to both $a_1$ and $a_3$. 
In the negotiation phase, if there are robots from $a_1$ and $a_3$ that want to migrate to $a_2$, there would be requests from $s_1$ to $s_2$, and from $s_2$ to $s_3$. 
$s_2$ (resp. $s_3$) will take care of border assignment between $a_1$ and $a_2$ (resp. $a_2$ and $a_3$). 
This could result in a collision at $c$ because $s_3$ would not know which borders $s_2$ has assigned for the robots in $a_1$ to come into $a_2$ -- the corner $c$ could be among them. 
The collision will happen if $s_3$ also assigns a robot in $a_3$ to come into $a_2$ at $c$. 
In the rejection phase, every active solver checks for this case and report to the neighboring solvers their robots that will cause the collision.
The reported solvers will keep the robots in their area for the current round and attempt to migrate them again in the next round, while the reporting solvers will treat them as not coming in (in this round).

\subsubsection{Movement Planning}
\label{sec:problem-solving-move}
A \emph{movement plan} is a sequence of movements for robots to reach their goal without collision. 
The program shown in figure~\ref{fig:prog-movement} is used by every active solver to make a movement plan for its area that exists some robot that has not reached its goal or assigned border yet. 
It is similar to the encoding used by the \emph{ASPRILO} project~\cite{gebser2018experimenting}, except that it has the following constraints: 

\begin{itemize}
\item \verb|:- r(R,N,0), o(N,_), not m(R,_,_,1), q(t).| -- forces incoming robots to move into the area at the first time step ($t = 1$).
\item \verb|:- q(t), g(t), r(R,N,t), c(N), not g(R,N).| -- prohibits robots to stay at any node that will be migrated into unless it is their goal.
\end{itemize}

The second constraint helps to increase the success rate of \emph{DMAPF}, but it could make movement planning unsuccessful unless there is sufficient free spaces in the area to enforce it.
Therefore, we only add atoms $c(N)$ when $(n_a - n_r - n_i) \geq n_f$, where, for an area $a$; $n_a$, $n_m$, $n_i$, and $n_f$ denote the number of nodes; the number of robots currently in it; the number of robots assigned to come in; and the minimum number of free nodes required, set by the user (currently $n_f = 4$), respectively. 
Most of the atoms come directly from $P_i$, except:

\begin{itemize}
\item $r(R,N,T)$ -- robot $R$ is at node $N$ at time step $T$. 
An atom $r(R,N_0,0)$ needs to be added for each robot $R$ in the area to denote its starting node $N_0$. 
\item $c(N)$ -- denotes that there will be a robot migrating into node $N$ in the next round. 
The set of such nodes is determined from the result of the \emph{negotiation} and the \emph{rejection} phases.
\end{itemize}

\begin{figure}
\begin{verbatim}
#program movement(t).
#external q(t).
{ m(R,D,N2,t) : x(N1,N2,D) } :- r(R,N1,t-1).
:- m(R,D1,_,t), m(R,D2,_,t), D1 != D2.
:- r(R,N,0), o(N,_), not m(R,_,_,1), q(t).
r(R,N2,t) :- r(R,N1,t-1), m(R,D,N2,t), x(N1,N2,D).
r(R,N,t) :- r(R,N,t-1), not m(R,_,_,t).
w(N1,N2,t) :- r(R,N1,t-1), m(R,D,N2,t), x(N1,N2,D).
:- w(N1,N2,t), w(N2,N1,t).
:- { r(_,N,t) } > 1, i(N,_).
g(t) :- r(R,N,t) : g(R,N).
:- q(t), not g(t).
:- q(t), g(t), r(R,N,t), c(N), not g(R,N).
#show m/4.
\end{verbatim}
\caption{The program to make a movement plan for multiple robots.}
\label{fig:prog-movement}
\end{figure}

Program \verb|movement(t)| is solved sequentially from $i = 0,\dots,h_m$ or until an answer set is found. 
$h_m$ is set to $(\sqrt{n_a} + 1) \cdot 2.0 \cdot F$, where $n_a$ is the number of nodes in area $a$, and $F$ is a sensitivity constant set by the user (currently $F = 2$). 
The lower the value of $F$ is, the less chance that a movement plan will be found, but it could terminate earlier if there is actually no plan, and vice versa. 
If no answer set is found after $i = h_m$, then we consider it to be impossible to make a movement plan for the given configuration, and \emph{DMAPF} will try to relax the configuration by removing the goal of a migrating robot one at a time, ordered by the Manhattan distance of the robot to its assigned border (longest first), and restarts the call to the program from $t = 0$. 
\emph{DMAPF} terminates (and returns no solution) if it cannot relax the configuration any further.

\subsubsection{Confirmation}
\label{sec:problem-solving-comm-3}
\emph{Confirmation} is the third, and the last phase of the migration process. 
After every active solver has finished movement planning, it will send the information of all the robots that are actually able to move out to the next area, in the neighboring solver, accordingly. 
The information of the robots remains mostly the same, except that their abstract plan will be shorten by one step because the current area will be removed as they have already been able to move out, and their current location is updated to the border node that they will be migrating from. 

\section{Experiment}
\label{sec:experiments}

Table~\ref{table:results-comparisons} provides comparison between three MAPF solvers: (\textit{i})~\emph{DMAPF}; (\textit{ii})~\emph{ros-dmapf}~\cite{pianpak2019distributed}; and (\textit{iii}) \emph{ECBS}\footnote{\url{https://github.com/whoenig/libMultiRobotPlanning}}~\cite{barer2014suboptimal} -- a relaxed version of \emph{CBS}~\cite{sharon2015conflict} where the solution quality is bounded-suboptimal instead of optimal. 
We also tried \emph{asprilo}~\cite{gebser2018experimenting} -- an ASP encoding that solves MAPF problems optimally -- but it cannot solve any of the problem instances within the timeout, so it has been excluded from the table. 
There are 15 problem configurations according to each row. 
Their map size is denoted in the first column (from the left). 
Each configuration contains a different number of robots according to the second column ($n_R$), and each robot is randomly assigned a different goal location. 
The problem instances do not include obstacles because we would be unable to compare with \emph{ros-dmapf} since it does not support having obstacles. 
The last three main columns contain results from the three solvers, which are \emph{Time} -- the solving time in seconds; \emph{Span} -- makespan of the solution; and \emph{Moves} -- the total number of moves from all the robots. 
The size of subproblems for both \emph{DMAPF} and \emph{ros-dmapf} is set to 8x8. 
The suboptimality bound for \emph{ECBS} is set to $1.2$. 
The tests were performed on Ubuntu 20.04 with \emph{Clingo} 5.5.0, running on a Dell Precision-3630 Tower with Intel i9-9900K CPU and 64 GB of RAM. 

\begin{table*}
\centering
\begin{tabular}{|c|c|c|c|c|c|c|c|c|c|c|}
\hline
\multirow{2}{*}{Map} & \multirow{2}{*}{$n_R$} & \multicolumn{3}{c|}{\emph{DMAPF}} & \multicolumn{3}{c|}{\emph{ros-dmapf}} & \multicolumn{3}{c|}{\emph{ECBS}} \\
\cline{3-11}
       &      & Time & Span & Moves  & Time   & Span & Moves  & Time   & Span & Moves \\
\hline
\multirow{5}{*}{\rotatebox[origin=c]{90}{24 x 24}} & 23   & 2.5  & 41   & 443    & 1.6    & 46   & 419    & 0.1    & 38   & 412   \\
       & 46   & 2.6  & 44   & 960    & 1.8    & 58   & 884    & 0.1    & 39   & 797   \\
       & 69   & 2.8  & 51   & 1432   & 2.0    & 69   & 1278   & 0.7    & 40   & 1132  \\
       & 92   & 3.6  & 57   & 2119   & 2.2    & 80   & 1883   & 1.4    & 40   & 1574  \\
       & 120  & 3.9  & 61   & 2751   & 2.5    & 82   & 2595   & 2.5    & 39   & 1973  \\
\hline
\multirow{5}{*}{\rotatebox[origin=c]{90}{48 x 48}} & 92   & 6.8  & 104  & 2890   & 3.8    & 118  & 2768   & 2.2    & 76   & 2651  \\
       & 184  & 9.8  & 117  & 6815   & 4.9    & 137  & 6625   & 11.8   & 81   & 5929  \\
       & 276  & 10.4 & 128  & 11683  & 7.7    & 187  & 11223  & 87.7   & 87   & 9387  \\
       & 368  & 13.2 & 124  & 16090  & 8.3    & 187  & 15858  & 191.7  & 86   & 12481 \\
       & 460  & 15.6 & 125  & 20920  & 11.0   & 210  & 20428  & -      & -    & -     \\
\hline
\multirow{5}{*}{\rotatebox[origin=c]{90}{96 x 96}} & 369   & 31.8  & 225 & 25041  & 30.0   & 242  & 24513  & -      & -    & -     \\
       & 737   & 41.7  & 240 & 52916  & -      & -    & -      & -      & -    & -     \\
       & 1106  & 61.8  & 280 & 88943  & -      & -    & -      & -      & -    & -     \\
       & 1474  & 83.2  & 282 & 124374 & -      & -    & -      & -      & -    & -     \\
       & 1843  & 175.4 & 282 & 165573 & -      & -    & -      & -      & -    & -     \\
\hline
\end{tabular}
\caption{Comparison between \emph{DMAPF}, \emph{ros-dmapf}, and \emph{ECBS}. The timeout is 180 seconds.}
\label{table:results-comparisons}
\end{table*}


\section{Conclusion and Discussion}
\label{sec:conclusion}
The experiment in Section~\ref{sec:experiments} shows that, while \emph{ECBS} performs very fast on the small problem instances, its performance becomes noticeably worse in a 48x48 map with 276 robots, and is unable to find the solution within the timeout in the bigger problems.  
\emph{ros-dmapf} is generally slightly faster than \emph{DMAPF}, but it is due to its simpler solver interaction protocol. 
It is unable to scale further when the maps get denser -- in terms of the ratio between the number of robots to the number of borders within an area. 
When this ratio increases, there is more chance for border assignments that will result in no movement plan to be selected.  
Since \emph{ros-dmapf} randomly assigns borders to migrating robots and cannot change the assignment afterwards, it has difficulties to scale further. 
The resulting makespan of \emph{DMAPF} is consistent, and better than \emph{ros-dmapf} due to the better border assignment mechanism, but it comes with a slight increase in the number of moves. 
The solution quality of \emph{ECBS} is still better than the other two solvers due to its centralized nature, but not by much in many cases. 
We believe \emph{DMAPF} can scale further given a better hardware and/or the use of clusters.  
To scale \emph{DMAPF} even further, the question whether it is feasible to also backtrack when movement planning fails even after all the relaxation should be addressed. 
Mechanisms that can balance the density of robots across the areas may also be helpful. 


\nocite{*}
\bibliographystyle{eptcs}
\bibliography{generic}
\end{document}